\documentclass[nato,noid,numreferences]{crckbked}

\usepackage{graphicx}
\usepackage{epsf}
\def\lab#1      {\hbox{\small #1} }
\newcommand{\be}{\begin{eqnarray}}
\newcommand{\ee}{\end{eqnarray}}
\newcommand{\ben}{\begin{eqnarray*}}
\newcommand{\een}{\end{eqnarray*}}
\newcommand{\la}{\langle}
\newcommand{\ra}{\rangle}
\newcommand{\half}{\frac{1}{2}}
\newcommand{\pe}{\rightarrow}

\def\mb#1         {\mbox{\boldmath $#1$}}
\def\diffn#1	  {\Delta^{-}_{#1}}

\newcommand{\AmS}{{\protect\the\textfont2
  A\kern-.1667em\lower.5ex\hbox{M}\kern-.125emS}}

\begin{document}

\begin{opening}
\title{Connections between Thin,Thick and Projection Vortices in SU(2) Lattice Gauge Theory}



\author{Richard W. Haymaker\thanks{\footnotesize Presented at the 
NATO Advanced Research Workshop
 on Confinement, Topology, and other Non-Perturbative 
 Aspects of QCD in Stara Lesna, Slovakia, Jan. 21-27, 
 2002.}}
\institute{Department of Physics and Astronomy\\ Louisiana State University\\
           Baton Rouge, Louisiana 70803 USA}

\author{Andrei Alexandru}
\institute{Department of Physics, University of Colorado\\
           Boulder, Colorado 80304 USA}



%

\begin{abstract}
We elucidate the connection between $ SO(3) \times Z(2)$ and the usual $SU(2)$ configuration variables.  
By exploiting the freedom of choosing a particular $SO(3)$ representative
we find a direct connection between
the two configuration spaces.  We are then able to compare the Kovacs-Tomboulis 
formulation of center vortices with the projection  vortex formulation on the same configuration.
Choosing a different representative, and going to the maximal center gauge, 
we show that projection vortices occur without approximation.  
The projection vortex dominance approximation results from dropping a factor in an exact
expression for the Wilson loop.
\end{abstract}

\end{opening}


\section{Introduction}

In the interest of finding a precise definition of thick center vortices on the lattice, 
Tomboulis\cite{t} and later Tomboulis and Kovacs\cite{kt1} reformulated SU(2) gauge theory in terms 
of variables defined on the factor groups $SO(3)$ and $Z(2)$.  The partition function and operators 
expressed in these variables are invariant under a sign flip of each link.
A particular choice of the sign of the trace of each link variable corresponds to the
choice of a {\em representative} of $SO(3)$. The bookkeeping that preserves the SU(2) theory 
is provided by new Z(2) valued plaquette variables  forming closed thin vortices.
A sign flip of a link forming a Wilson loops is accompanied 
by the introduction of a vortex linked to it 
giving a compensating tiling factor  of $-1$.

In a recent paper\cite{ah3} we found a  direct kinematical connection  between this approach and the 
usual SU(2) formalism. For a particular $SO(3)$ representative 
the two formulations are identical.  Therefore we can define
both the Kovacs-Tomboulis\cite{kt1} (KT) and the Projection (P) vortex counters\cite{dfgo} 
on the same configuration giving a direct comparison of these two approaches.    

Secondly, by choosing another  $SO(3)$ representative 
we showed that the resulting thin vortices are precisely the structures known as 
P vortices.  The projection
approximation comes from a subsequent truncation.

Although the $SO(3) \times Z(2)$ formulation is equivalent to the $SU(2)$ form, 
the former is particularly useful in illustrating connections between 
the KT and P vortex approaches.  Configurations are generated
most efficiently in the SU(2) variables.  One can  match the $SU(2)$ variables to the corresponding 
$SO(3) \times Z(2)$ variables.  Then we are free to flip the signs of the links thereby
changing representative.  If one chooses signs for which the trace of all links are positive, 
then the thin vortices created in the process will be identical to the those generaged by the 
projection algorithm\cite{dfgo}.  

In the $SO(3) \times Z(2)$  formulation, Wilson loops have a perimeter factor and a tiling factor.  
Changing the representative may move a  minus sign from one factor to another, but  always leaves
the product invariant.   This separation  
is gauge dependent.  The goal of the projection approach is to transfer the disordering
from the perimeter factor to the tiling factor since the 
the projection approximation consists of setting the perimeter factor to one.  
The quality of the 
P vortex approximation then depends on one's ability to suppress the disordering
mechanism in the perimeter factor through a judicious choice
of gauge\cite{dfggo,kt2,born1,bfgo,born2,fgo}.

\section{$SU(2)$ configurations in $SO(3) \times Z(2)$ variables}

The Wilson form of the partition function can be recast by introducing
$Z(2)$ valued independent variables $\sigma(p)$ defined on 
plaquettes\cite{t,ah1}
\ben
Z_{SO(3) \times Z(2)}  &=& 
\int 
\left[
dU(b) 
\right]
\sum_{ \sigma(p)}
\\
&&
\left[
\prod_c
   \delta\left(\sigma(\partial c) \eta(\partial c)\right)
\right]
\\
&&
\exp
\left(
   \beta \sum_{p} \half|Tr[  U(\partial p)]| \sigma(p)
\right),
\een
where the dependent variables $\eta(p)$ are defined by
\ben
Tr [U(\partial p)] \equiv |Tr [U(\partial p)] | \eta(p).
\een
The ``cube constraint" factor requires that  $\prod_1^6 \eta(p)\sigma(p) = +1$ over the 
six faces of all cubes.

Wilson loops have  $Z(2)$ valued plaquette tiling factors, $\sigma$ and $\eta$ on an 
arbitrary surface $S$ bounded by $C$
\be
W_{m \times n}(C) &=& \lab{Tr} [  U(C)]  \eta(S) \sigma(S)|_{C = \partial S}, 
\label{wilson}
\ee
\ben
W_{1 \times 1} &=&  \lab{Tr} [  U(\partial p)] \eta(p)  \sigma(p) 
= |\lab{Tr} [  U(\partial p)] |  \sigma(p).
\een

Properties of this form include:
\begin{itemize} 

   \item$Z(2)$ invariance of $Z$ and of observables under   $U(b) \rightarrow - U(b)$. 
    There are therefore $2^N$ {\em representatives} of $SO(3)$, where $N$ is the number of links. \\

   \item There exist  co-closed $\sigma(p) - \eta(p)$ vortex sheets due to the cube constraint with 
        patches of either 
         $\sigma(p)=-1$ or $\eta(p)=-1$, $\sigma(p) \eta(p) = -1$. 
      Pure $\sigma(p)$ or $\eta(p)$  vortex sheets are limiting cases.\\
   \item A change of representative can deform existing $\eta$ patches and create or destroy pure
     $\eta$ vortex sheets.

\end{itemize}

\subsection{The representative $\widetilde{U}$}

This is defined by the condition
\ben
\sigma(p)\eta(p) &=& +1,\;\;\;\forall p.
\een
In this case the cube constraint is automatically satisfied.
There are further simplifications:
\ben
|Tr[  \widetilde{U}(\partial p)]| \sigma(p) 
&=& Tr[  \widetilde{U}(\partial p)] \eta(p) \sigma(p),
             \\ &=& Tr[  \widetilde{U}(\partial p)], 
\een
\ben
\widetilde{Z}
&=& 
\int 
\left[
    d \widetilde{U}(b)
\right]
\exp
\left(
   \beta 
   \sum_{p} 
   \half\lab{Tr} [  \widetilde{U}(\partial p)] 
\right),
\een
\ben
W_{m \times n} = \lab{Tr} [  \widetilde{U}(C)] , \;\; 
  W_{1 \times 1} =  \lab{Tr} [  \widetilde{U}(\partial p)].
\een

We showed\cite{ah1,ah3} that starting from a cold configuration,
$U(b)=\sigma(p)=+1$, we can reach the full configuration space of the 
independent variables $\{U(b), \sigma(p)\}$ through local updates 
while staying in the representative $\widetilde{U}(b)$.  
In this representative all $\sigma-\eta$ vortices are absent.

This particular representative provides the connection of this formulation to the 
SU(2) formalism with the Wilson action.
As a consequence, we can define the Tomboulis thin, thick and hybrid
vortex counters on ordinary $SU(2)$ configurations as will be given below.

\subsection{The representative $\widehat{U}$}

This is defined by the condition 
\ben
\lab{Tr} [\widehat{U}(b)] &\ge& 0.
\een
This can be obtained by a single sweep.
The interest in this is to connect with P vortices which are defined 
as follows:  One first fixes the gauge, for example the maximal center gauge and then

{\bf In an arbitrary representative}
\begin{itemize} 
      \item Project: $ sign \lab{Tr} [U(b)] \rightarrow u(b)$, $u(b) = \pm 1$.
      \item $P$ vortex: $u(p) = u(\partial p) \eta(p) \sigma(p)=-1$
      \item Proj. approx.:  $W(C) \approx   u(S)|_{C = \partial S}$.
\end{itemize}

{\bf In the  $\widehat{U}(b)$ representative}
\begin{itemize} 
      \item Project: $ sign \lab{Tr} [\widehat{U}(b)] \rightarrow \widehat{u}(b)$,
       $\widehat{u}(b) = + 1$.
      \item $P$ vortex: $\widehat{u}(p) = \eta(p) \sigma(p)=-1$, which is identical to
$\sigma - \eta$ vortex.
        \item Proj. approx.:   $\lab{Tr} [\widehat{U}(C)] \approx 1 $,
\end{itemize}
where we have used Eqn.(\ref{wilson}).  These two procedures give identical P vortices. 

However in the $\widehat{U}(b)$ representative
the $P$ vortices are identical to the $\sigma - \eta$ vortices which are a
tiling factor in the exact definition of the Wilson loop.  The success or failure of a projection approximation 
depends on whether  one can find a gauge such that the sign fluctuations of the perimeter 
factor in Eqn.(\ref{wilson}) can be transferred to the tiling factors arising from $\sigma - \eta$ linkages.
If so then one argues that the area law of a Wilson loop arises
from P vortex linkages in that gauge.

\section{Kovacs-Tomboulis vortex counters}
Kovacs and Tomboulis\cite{kt1} gave representative independent definitions of three
vortex counters based on  $SO(3) \times Z(2)$  configurations.
\ben
N_{thin}(S)&=&\prod_{p\in S} \sigma(p), \\
N_{thick}(S)&=&\lab{sign} \; \lab{tr} [ U(C) ] \times \prod_{p\in S} \eta(p), \\
N_{hybrid}\;\;\;\;&=& N_{thin}(S) \times N_{thick}(S) = \lab{sign} \, W.
\een
The hybrid counter is necessarily independent of surface. 
$N_{thin}(S)$ and $N_{thick}(S)$ count the corresponding vortices only if the
value is independent of surface $S$.

We can express these counters in terms of $SU(2)$ configurations by evaluating the
above expressions in the $\widetilde{U}(b)$ representative.
\ben
N_{thin}(S)&=&\prod_{p\in S} \lab{tr} [\partial \widetilde{U}(p)], \\
N_{hybrid}\;\;\;\;&=& \lab{sign} \,\, \lab{tr} [\widetilde{U}(C)], \\
N_{thick}(S)&=& \prod_{p\in S} \lab{sign} \,\,\lab{tr} [\partial \widetilde{U}(p)]\times 
\lab{sign} \,\, \lab{tr} [\widetilde{U}(C)].
\een

\section{Numerical Results}

It is not feasible to measure these counters on all possible surfaces.  We made measurements only on the minimal
surface\cite{ah3,ah1}.  As a consequence, a measurement giving for example $N_{thin}(S) = -1$ indicates only the occurrence
of an odd number of $\sigma$ patches which could be part of  thin or hybrid vortices. And similarly for
the thick case.

The contribution to the potential from the three types of vortex counters is
\ben
V(R)=-\lim_{T\pe\infty} \frac{1}{T} \ln \la N(W(R, T)) \ra,
\een
where $N(W(R,T))$ is the thin, thick or hybrid counter signal for that particular Wilson loop (taking values  $\pm 1$). 

Fig. 1 shows that the string tension in $V_{thin}$ in 
physical units {\em increases} in the approach to the continuum limit.
Although this is perhaps surprising, we showed that
this is canceled by an increasing string tension in the thick potential\cite{ah3}.

The K-T definition for vortices\cite{kt1} is appealing since it is  gauge invariant 
 but they are hard to localize on a lattice. 
P vortices\cite{dfgo}, on the other hand, are easy to localize but are not gauge invariant. 
It is interesting to see if these two definitions agree. 
We now have the tools to compare these definitions of  vortex counters 
on the same configuration. 

Fig. 2 shows plots of the average of the fraction odd/(odd+even)  hybrid and P vortices linking a Wilson loop as 
a function of area. The average of the product compared to the product of the average shows that there is essentially no correlation.  The corresponding plots for thin vortex fractions and thick ones gives essentially the same
result.  In Ref.\cite{ah3} we examine more sensitive signals of correlations but without a definitive result.

\begin{figure}
\centering
\includegraphics[width=8cm]{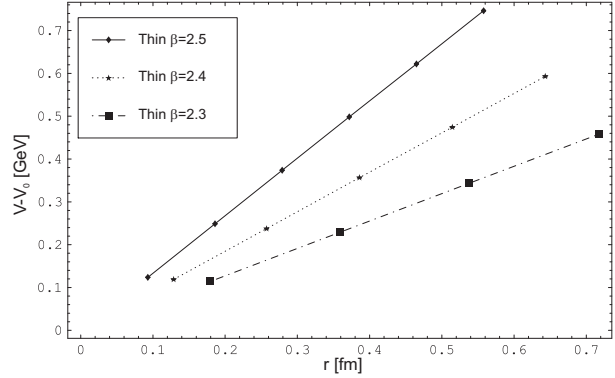}
\caption{Vortex potentials in physical units for the thin counter.}
\label{Fig1}
\end{figure}

\begin{figure}
\centering
\includegraphics[width=8cm]{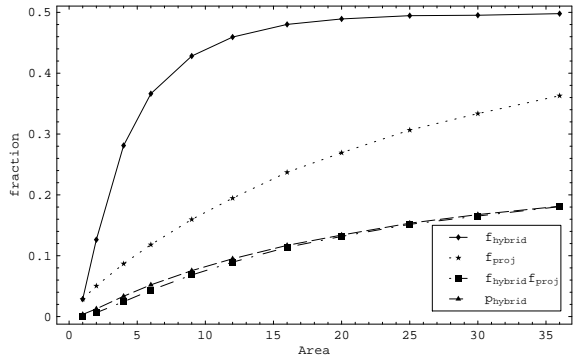}
\caption{Average of the fraction  odd/ (odd+even) hybrid and projection vortices linking a Wilson loop as 
a function of area.}
\end{figure}

\section{Idealized vortex configurations}

As we see from the numerical results it is problematic to assign the same physics to the 
KT and P  vortex counters because of the absence of a simple correlation. 
To understand the reasons for this it may be a useful exercise to look at an idealized configuration for which 
the two approaches give the same signal. We
are thinking of the situation in which center vortices are well defined and isolated from one another
by regions of pure gauge. 
We denote  the domain of pure gauge by  ${\cal D}$ for which all links are gauge 
equivalent to $\pm I$ and a complementary domain in which the field strength may be  non-zero.  
Assume in general that ${\cal D}$ is multiply connected.  

Go to the gauge where all links in ${\cal D}$  $= \pm I$. 
The advantage of the $SO(3) \times Z(2)$  variables is that we can subsequently  go to the 
representative such that all links in ${\cal D}$ $ = + I$.  Consider a Wilson loop in this
multi-connected region ${\cal D}$.  The perimeter factor $=+1$.  The tiling factors
are due to the thin vortices linked to the loop.

\begin{description}
\item [KT interpretation] Center vortices are defined only by their topological linkage.  The linkage
 is unambiguous only if the perimeter of the 
loop is completely in the domain ${\cal D}$. The value of the loop in this
case is
determined by the tiling factor alone which counts thin $\eta-\sigma$ vortex linkages, $\lab{mod} (2)$.  
If there are $\sigma$ patches present in the $\eta-\sigma$ vortex, then
the thick vortex degenerates to a thin vortex at those locations but in general spreads out forming
a thick vortex elsewhere.  KT refers to this as a hybrid vortex.

\item [P interpretation] The construction has fixed the positions of the P vortices. 
One can locate and count those linking the loop giving a factor of $(-1)^n$ for the Wilson 
loop for the case of $n$ vortices.

\item[Comparison] The two approaches arrive at the same value for the Wilson loop on this idealized configuration. 
The P and $\eta-\sigma$ vortices are the same. 
Both approaches come to the same conclusion on the presence or absence of 
a bona fide center vortex, mod($2$).  
However the KT definition is gauge invariant and representative invariant and therefore any particular
details in this gauge and this representative such as location of the thin vortices is not particularly
relevant. The P vortices are fixed by the construction and the pattern  could indicate a substructure.  However
identifying substructure center vortices can not be tested  by 
the KT topological definition if it involves a 
Wilson loop perimeter that strays from the domain ${\cal D}$.

\end{description}

If this simple picture has some validity, the numerical results suggests that it is obscured 
by noise in the domain ${\cal D}$ and/or vortex cores that overlap, among many other possibilities.
It might be helpful if one could find a related theory in which the dynamics creates vortices on one time
scale establishing their identity and the vortices move and deform on longer time scale.

\section{Summary and Conclusions} 
We have shown that
\begin{itemize} 
\item An $SO(3) \times Z(2)$ configuration is identical to an $SU(2)$ configuration in a particular
representative.  Updates of the former can be done with the simpler $SU(2)$ variables.
\item
The $\eta-\sigma$ vortices of Kovacs and Tomboulis are identical to P vortices in an
particular gauge and a particular representative.
\item The KT vortex counters are gauge invariant and representative invariant and are measurable
on $SU(2)$ configurations.
\item 
The string tension due to $\sigma$ patches of thin vortices taken alone 
has a surprising and definitive signal of increasing string tension as
$a \rightarrow 0$.  The thick patches behave similarly and taken together, the 
scaling violations  cancel.
\item
A simple test for correlations of KT and P vortex counters gives a null result.  
More sensitive tests have been reported elsewhere\cite{ah3} but without definitive results. 
\end{itemize}

\section{Acknowledgments} 
This work was supported in part by the United States Department of Energy, grant DE-FG05-91 ER 40617.

\end{document}